\documentclass[10pt]{article}
\usepackage{amssymb}
\usepackage{bm}
\usepackage{amsmath}
\usepackage{amscd}
\newcommand{\ket}[1]{ | #1  \rangle}
\newcommand{\bra}[1]{ \langle #1  |}

\def\openone{\leavevmode\hbox{\small1\kern-3.8pt\normalsize1}}
\def\tr{{\rm tr} }

\def\ch{{\cal H}}

\def\ct{{\cal T}}
\def\cs{{\cal S}}

\def\RR{\mathbb{R}}

\newtheorem{theorem}{Theorem}

\newtheorem{lemma}{Lemma}

\newtheorem{property}{Property}

\newtheorem{corollary}{Corollary}

\newcommand{\ketbra}[1]{{\ket{#1}\!\bra{#1}}}

\newcommand{\proj}[1]{\ket{#1}\!\bra{#1}}

\newcommand{\inner}[2]{ \langle #1 | #2 \rangle}

\newcommand{\1}{{\openone}}

\newcommand{\beq}{\begin{equation}}
\newcommand{\eeq}{\end{equation}}
\newcommand{\beqa}{\begin{eqnarray}}
\newcommand{\eeqa}{\end{eqnarray}}

\newcommand{\qed}{\Box}

\begin{document}
\begin{center}
{\Large\bf On the existence of physical transformations\\ between sets of quantum states }\\
\bigskip
{\normalsize  Anthony Chefles$^\S$, Richard Jozsa$^\dagger$ and Andreas Winter$^\dagger$}\\
\bigskip

{\small\it
$^\S$Department of Physical Sciences, University of Hertfordshire,\\
Hatfield AL10 9AB, Herts, U.K.\\[1mm]

$^\dagger$Department of Computer Science, University of Bristol,\\
Merchant Venturers Building, Bristol BS8 1UB U.K. }
\\[4mm]
\date{today}
\end{center}

\begin{abstract} Let $A=\{ \rho_1, \ldots ,\rho_n \}$ be a given
set of quantum states. We consider the problem of finding
necessary and sufficient conditions on another set $B=\{
\sigma_1,\ldots ,\sigma_n \}$ that guarantee the existence of a
physical transformation taking $\rho_i$ to $\sigma_i$ for all $i$.
Uhlmann has given an elegant such condition when both sets
comprise pure states. We give a simple proof of this condition and
develop some consequences. Then we consider multi-probabilistic
transformations between sets of pure states which leads to
conditions for the problem of transformability between $A$ and $B$
when one set is pure and the other is arbitrary.

\end{abstract}

%newpage

\section{Introduction}\label{intro}
Quantum information theory is often concerned with the
manipulation of {\em families} of quantum states rather than
individual states in isolation. For example a discrete quantum
source is defined by a (finite) family of signal states together
with prior probabilities. In this paper we will address the
following fundamental question of transformability in the context
of families of states. Let $\{ \rho_1, \ldots ,\rho_n \}$ be a
given (finite) set of quantum states. What are necessary and
sufficient conditions on another set $\{ \sigma_1, \ldots
,\sigma_n \}$ that guarantee the existence of a physical
transformation $\ct$ satisfying $\ct (\rho_i)=\sigma_i$ for all
$i$? We will write $\{ \rho_1, \ldots ,\rho_n \}\Rightarrow \{
\sigma_1,\ldots ,\sigma_n \} $ to denote the statement that such a
physical transformation exists. We will call $\rho_i$ the source
states and $\sigma_i$ the target states.

By a physical transformation $\ct$ we mean a linear completely
positive trace preserving map on density operators. In standard
quantum formalism this is the most general change that a state can
undergo by any physical process. Let $\rho^X$ be any
(finite-dimensional) density operator of a system $X$. We have two
well known mathematical expressions for physical
transformations.\\ (a) Kraus operator sum form: \[ \ct
(\rho^X)=\sum_\mu A_\mu \rho^X A_\mu^\dagger \] where $\{ A_\mu
\}$ is any set of operators satisfying $\sum_\mu A_\mu^\dagger
A_\mu =I$ (and $^\dagger$ denotes the adjoint).\\ (b) Stinespring
unitary dilation form: introduce an auxiliary (ancilla) system $E$
in standard state $\proj{0}$. Then \[ \ct (\rho^X)= \tr_E\, U^{XE}
(\rho^X\otimes \proj{0}^E)U^{XE \dagger} \] where $U_{XE}$ is a
unitary operation on the joint system $XE$.

$F(\rho,\omega)$ will denote the fidelity between states $\rho$
and $\sigma$, defined by \[ F(\rho,\omega)=  \tr
\sqrt{\sqrt{\omega}\rho\sqrt{\omega}}. \] We also have
\cite{uhlmann,J94}
\begin{equation} \label{purifs} F(\rho, \omega)=\max
|\inner{\xi_\rho}{\xi_\omega}| \end{equation} where the maximum is
taken over all choices of purifications $\ket{\xi_\rho}$ of $\rho$
and $\ket{\xi_\omega}$ of $\omega$.

We begin by reviewing some known results about the
transformability problem. This question (and its restriction to
some special cases such as commutative states) has been considered
in the work of Alberti and Uhlmann \cite{alb-uhl0,AE} since the
early 1980s. For $n=1$ the problem is trivial ($\rho\Rightarrow
\sigma$ for any $\rho$ and $\sigma$) but already for $n=2$ (pairs
of states) the general problem is difficult and mostly unsolved.
The significance of fidelity for the transformability problem is
summarised in lemma \ref{lemma1}. For pairs of states this
question was studied in \cite{uhl85.2}.

\begin{lemma} \label{lemma1}(a) Suppose that $\{ \rho_1,\ldots ,\rho_n
\}\Rightarrow \{ \sigma_1, \ldots ,\sigma_n \}$. Then
$F(\sigma_i,\sigma_j)\geq F(\rho_i,\rho_j)$ for all $1\leq i,j\leq
n$.\\ (b) Suppose that $\rho_1$ and $\rho_2$ are {\em pure}
states. Then $\{ \rho_1, \rho_2 \} \Rightarrow \{
\sigma_1,\sigma_2 \}$ iff $F(\sigma_1,\sigma_2)\geq
F(\rho_1,\rho_2)$. \end{lemma} We omit the proof (which may be
readily seen from the Stinespring form of a physical
transformation and eq. (\ref{purifs})).

Thus the fidelity conditions $F(\sigma_i,\sigma_j)\geq
F(\rho_i,\rho_j)$ are always necessary. They are also sufficient
in the case of pairs so long as the source states are also pure
(but fail to be sufficient for general source pairs even if the
targets are pure). For triples of states these conditions also
fail to be sufficient, even if all source as well as all target
states are pure. (Counterexamples may be found in \cite{JS99}).

In 1980 Alberti and Uhlmann \cite{AE} found necessary and
sufficient conditions for the transformability of general {\em
pairs} of {\em qubit} states. Let $||A||_1 = \tr
\sqrt{AA^\dagger}$ denote the trace norm of an operator $A$. If
$A$ is hermitian then $||A||_1=\sum |\lambda_i|$ where $\lambda_i$
are the eigenvalues of $A$.

\begin{theorem}\label{aethm}\cite{AE} Suppose that
$\rho_1,\rho_2,\sigma_1,\sigma_2$ are all qubit states. Then
$\{\rho_1,\rho_2 \}\Rightarrow \{\sigma_1,\sigma_2 \}$ iff
\begin{equation}\label{conds} ||\sigma_1-t\sigma_2||_1 \leq
||\rho_1-t\rho_2||_1 \hspace{5mm} \mbox{for all $t\in \RR^+$}.
\end{equation}
\end{theorem}
It may be shown \cite{papriv} that eq. (\ref{conds}) is necessary
in any dimension but fails to be sufficient already in dimension
3. However if $\{\rho_1,\rho_2 \}$ and $\{ \sigma_1,\sigma_2 \}$
are both commuting pairs of density operators then eq.
(\ref{conds}) is necessary and sufficient in any dimension. Some
applications of theorem \ref{aethm} have been given in
\cite{sasaki}.

By a simple rescaling, eq. (\ref{conds}) may be written
equivalently as \[ ||p_1\sigma_1-p_2\sigma_2||_1 \leq
||p_1\rho_1-p_2\rho_2||_1 \hspace{5mm} \mbox{for all probability
distributions $\{ p_1,p_2 \}$}. \] Then it is interesting to
recall Helstrom's minimum error probability $P_{E}^{\min }$
\cite{helstrom} for distinguishing a pair of quantum states
$\tau_1, \tau_2$:
\[ P_{E}^{\min }(\tau_1,\tau_2) =
\frac{1}{2}\left( 1-||p_1\tau_1-p_2\tau_2 ||_1 \right) \] Hence
theorem \ref{aethm} may be interpreted as saying that a physical
transformation exists for pairs of qubit states iff the target
states are no more distinguishable than the source states by
minimum error probability discrimination, for any prior
probabilities.

In \cite{alb03} the transformability problem for pairs was further
studied leading to an interesting set of necessary and sufficient
conditions, but involving a quantification over sets of comparable
complexity to the set of physical transformations itself.

In \cite{uhl85.1} Uhlmann considered the transformability problem
for sets of pure states, giving an elegant necessary and
sufficient condition, which was later independently rediscovered
by Chefles \cite{chefles1}.

In the present paper we begin by considering pure state
transformations, giving a simple proof of Uhlmann's condition
(theorem \ref{pure}) that improves on the argument in
\cite{chefles1}, and we develop some consequences (corollaries
\ref{cor1},\ref{cor2},\ref{cor3}). Then we go on to consider
so-called {\em multi-probabilistic} transformations between sets
of pure states, generalising a notion introduced in
\cite{chefles2}. We will see that our characterisation  of such
transformations (theorem \ref{thm:mp:st}) provides a set of
necessary and sufficient conditions  for the more general problem
of transformability between sets of states $\{\rho_1,\ldots
,\rho_n \}$ and $\{ \sigma_1,\ldots ,\sigma_n\}$, where just one
set is pure and the other is arbitrary (theorem \ref{puremix}).

\section{Sets of pure states}\label{purestates}
For any two matrices $C$ and $D$ of the same size, we introduce
the Hadamard product $C\circ D$ defined by entry-wise product
$[C\circ D]_{ij}=[C]_{ij}[D]_{ij}$ with no summing. We will also
write $M\geq 0$ to denote the statement that a matrix $M$ is
positive semidefinite.

Let $S_A=\{ \alpha_1, \ldots ,\alpha_n\}$ be a set of pure states
$\alpha_i=\proj{a_i}$. Given $\alpha_i$ the ket $\ket{a_i}$ is
fixed only up to a phase freedom. For any choice $K_A=\{
\ket{a_1},\ldots ,\ket{a_n}\}$ of such kets the Gram matrix $G_A$
is given by $[G_A]_{ij}=\inner{a_i}{a_j}$. Hence $G_A$ is not
determined uniquely by $S_A$ but we have the following.

\begin{lemma} \label{gramlemma} Let $G_A$ be any Gram matrix for
$S_A$. Then $G_A'$ is a Gram matrix for $S_A$ iff $G_A'=G_A \circ
K$ where $[K]_{ij}=e^{i(\theta_i-\theta_j)}$ for some
$\theta_1,\ldots ,\theta_n \in \RR$ (so $K\geq 0 $ too).
\end{lemma}

\noindent{\bf Proof}\,\, Easy to check. $\qed$

Thus we associate a whole family of Gram matrices (related as in
lemma \ref{gramlemma}) to a set $S_A$ of pure physical states, but
we could also attempt to specify a canonical unique choice. For
example suppose that $\alpha_i\alpha_j \neq 0$ for all $i,j$ i.e.
the states are pairwise non-orthogonal. Let $\ket{a_1}$ be any
choice of ket for $\alpha_1$ and then choose $\ket{a_i}$ for $i>1$
such that $\inner{a_1}{a_i}$ is real positive, which fixes these
kets uniquely. The resulting Gram matrix $G_{ij}=\inner{a_i}{a_j}$
is then independent of the choice of $\ket{a_1}$. Indeed directly
in terms of the states $\alpha_i$ we can write
\begin{equation}\label{gram1} \mbox{$G_{ij}=
r_{ij}e^{i\theta_{ij}}$ where $r_{ij}^2=\tr\, \alpha_i\alpha_j$
and $\theta_{ij}=$ phase of $\tr\, \alpha_1\alpha_i\alpha_j$}
\end{equation} (noting that the phase of $\tr\,
\alpha_1\alpha_i\alpha_j$ equals the phase of $\inner{a_i}{a_j}$
if $\inner{a_1}{a_i}$ and $\inner{a_1}{a_j}$ are real positive).
This construction, as written, fails to specify $G$ uniquely if
$\alpha_1\alpha_i=0$ for some $i$ and then presumably some more
complicated prescription is required. However in the following, we
will not work with such canonical choices.

We have the following three basic properties of Gram matrices.
\begin{property} \label{prop1} If $S_{AX}= \{ \alpha_i\otimes
\xi_i \}$ is a set of pure product states and $G_A$, $G_X$ are
Gram matrices for $\{\alpha_i\}$ and $\{ \xi_i \}$ respectively
then $G_{AX}=G_A\circ G_X$ (where the kets for $S_{AX}$ are taken
as the products of the chosen kets for $G_A$ and $G_X$).
\end{property}

\noindent {\bf Proof}\,\, Easy to check. $\qed$

\begin{property}\label{prop2} A matrix $M$ arises as a Gram matrix for
some set of kets $\{\ket{a_i}\}$ iff (i) $M_{ii}=1$ for all $i$,
 and (ii) $M\geq 0$.
\end{property}

\noindent {\bf Proof}\,\, The forward implication is immediate.
Conversely if $M\geq 0$ then we can write $M=C^\dagger C$. Then
the $\ket{a_i}$'s are given in components by the columns of $C$.
$\qed$

\begin{property}\label{prop3} Let $S_A=\{
\alpha_i\}$ and $S_B=\{ \beta_i \}$ be two sets of pure states and
let $G_A$ and $G_B$ be any corresponding choices of Gram matrices.
Then $S_A$ and $S_B$ are unitarily equivalent iff $G_A=K\circ G_B$
where $[K]_{ij}=e^{i(\theta_i-\theta_j)}$ for some $\theta_i\in
\RR$.\end{property}

\noindent {\bf Proof}\,\, By lemma \ref{gramlemma} we can choose
equal Gram matrices for $S_A$ and $S_B$ iff $G_A=K\circ G_B$ with
$[K]_{ij}=e^{i(\theta_i-\theta_j)}$ for some $\theta_i\in \RR$.
The result then follows immediately from lemma 1 of \cite{JS99},
stating that two sets of kets are unitarily equivalent iff they
have equal Gram matrices. $\qed$

Using these properties, we get the following result.

\begin{theorem}\label{pure}\cite{uhl85.1,chefles1}
 Let $\{\alpha_i\}$ and $\{ \beta_i\}$
be two sets of pure states. Let $G_A$ and $G_B$ be any
corresponding choices of Gram matrices. Then $\{ \alpha_i\}
\Rightarrow \{\beta_i\}$ iff $G_A=M\circ G_B$ for some $M\geq 0$
(and then we must also have $M$ satisfying (i) of property
\ref{prop2}).
\end{theorem}

\noindent {\bf Proof}\,\, There is a physical transformation
mapping $\alpha_i$ to $\beta_i$ \\ {\em iff} there is a unitary
map $\alpha_i\otimes \proj{0} \rightarrow \beta_i\otimes \xi_i$
for some pure states $\xi_i$ (by Stinespring dilation) \\
{\em iff} $\{ \alpha_i\}$ and $\{ \beta_i\otimes \xi_i\}$ are
unitarily equivalent for some choice of pure states $\xi_i$\\ {\em
iff} $G_A=K\circ G_X\circ G_B$ for some $G_X\geq 0$ and
$[K]_{ij}=e^{i(\theta_i-\theta_j)}$ for some $\theta_i\in \RR$ (by
properties \ref{prop1},\ref{prop3})\\ {\em iff} $G_A=M\circ G_B$
for some $M\geq 0$ (by property \ref{prop2}). \\ In the last step
we have also used the fact that the Hadamard product of positive
semi-definite matrices is positive semi-definite. $\qed$

\begin{corollary}\label{cor1} Let $\{ \alpha_i\}$ and $\{
\beta_i\}$ be sets of pure states on $\ch_A$ and $\ch_B$
respectively. \\ (a) If $\{ \alpha_i\}\Rightarrow \{ \beta_i\}$
and  $\{ \beta_i\}\Rightarrow \{ \alpha_i\}$ then $\{\alpha_i\}$
and $\{\beta_i\}$ are unitarily equivalent (i.e. there is a
unitary operation $U$ on $\ch_A\oplus \ch_B$ with
$U\alpha_iU^\dagger = \beta_i$ for each $i$).\\ (b) Let $\pi$ be
any permutation of the set of indices $i$. If
$\{\alpha_i\}\Rightarrow \{\beta_i\}$ and $\{\beta_i\}\Rightarrow
\{\alpha_{\pi(i)}\}$ then $\{\alpha_i\}$ and $\{\beta_i\}$ are
unitarily equivalent.\end{corollary}

\noindent {\bf Proof}\,\, (a) By theorem \ref{pure} we can write
\begin{equation}\label{star} \mbox{ $G_A=M^{(1)}\circ G_B$ and
$G_B=M^{(2)}\circ G_A$.} \end{equation} Let $K_A=\{\ket{a_i}\}$
and $K_B=\{ \ket{b_i}\}$ be sets of kets with Gram matrices $G_A$
and $G_B$ respectively. By eq. (\ref{star}), $[G_A]_{ij}=0$ iff
$[G_B]_{ij}=0$. Hence if the kets of $K_A$ lie in a family of
orthogonal subspaces then the corresponding kets of $K_B$ must
similarly fall into subsets that are mutually orthogonal, and the
corresponding subsets of $K_A$ and $K_B$ must be mapped to
eachother. So without loss of generality we may assume that $K_A$
cannot be orthogonally decomposed and in particular, for each
$\ket{a_k}$ there is a sequence from $\ket{a_1}$ to $\ket{a_k}$:
\begin{equation}\label{seq} \ket{a_1}=\ket{a_{l_0}},\ket{a_{l_1}},
\ldots , \ket{a_{l_m}}=\ket{a_k} \end{equation} such that each
successive pair $\ket{a_{l_i}}$ and $\ket{a_{l_{i+1}}}$ is
non-orthogonal. Now $G_A=M^{(1)}\circ M^{(2)}\circ G_A$ so if
$[G_A]_{ij}=\inner{a_i}{a_j}\neq 0$ then
\begin{equation}\label{inv}
[M^{(1)}]_{ij}=\frac{1}{[M^{(2)}]_{ij}}. \end{equation} But
$M^{(1)}$ is the Gram matrix $G_X$ for some set $\{ \ket{\xi_i}\}$
 and similarly for $M^{(2)}$. Hence $|[M^{(k)}]_{ij}|\leq 1$ and
 then eq. (\ref{inv}) gives $|[M^{(1)}]_{ij}|=1$ if
 $\inner{a_i}{a_j}\neq 0$. In that case
 $\ket{\xi_j}=e^{i\theta_{ij}}\ket{\xi_i}$. Now according to eq.
 (\ref{seq}) any $\ket{a_k}$ can be connected back to $\ket{a_1}$
 by a sequence of $\ket{a_i}$'s such that any two consecutive
 states are non-orthogonal. Hence the corresponding $\ket{\xi_i}$'s
 along the sequence differ only by a phase. Thus we get
 $\ket{\xi_k}=e^{i\theta_k}\ket{\xi_1}$ for all $k$ so
 $[M^{(1)}]_{ij}= e^{i(\theta_i-\theta_j)}$ and by property
 \ref{prop3}, $\{\alpha_i\}$ and
 $\{\beta_i\}$ are unitarily equivalent.\\ (b) Suppose $\ct
 :\alpha_i\rightarrow \beta_i$ and $\cs : \beta_i\rightarrow
 \alpha_{\pi(i)}$ are physical transformations. Then
 $\cs\ct:\alpha_i\rightarrow \alpha_{\pi(i)}$ and there is a power
 $k$ (actually the order of $\pi$ in the permutation group) such
 that $(\cs\ct)^k:\alpha_i\rightarrow \alpha_i$. Writing
 $(\cs\ct)^k$ as $((\cs\ct)^{k-1}\cs)\, \ct$ we see that
 $((\cs\ct)^{k-1}\cs):\beta_i\rightarrow \alpha_i$ so by (a),
 $\{\alpha_i\}$ and $\{\beta_i\}$ are unitarily equivalent. $\qed$

The method of proof of theorem \ref{pure} also allows us to
characterise the space of all possible physical transformations
taking $\{\alpha_i\}$ to $\{\beta_i\}$ when
$\{\alpha_i\}\Rightarrow\{\beta_i\}$ is true. For example we have
the following.

\begin{corollary}\label{cor2}
Let $\{\alpha_i\}$ and $\{\beta_i\}$ be sets of pure states.
Suppose that $\{\alpha_i\}\Rightarrow\{\beta_i\}$ and $\tr\,
\beta_i\beta_j \neq 0$ for all $i,j$ i.e. the pure target states
are pairwise non-orthogonal. Then there is a unique physical
transformation (for density operators on the joint support of the
source states) mapping $\{\alpha_i\}$ to $\{\beta_i\}$.
\end{corollary}

\noindent {\bf Proof}\,\, Suppose that we have a physical
transformation $\ct:\alpha_i\rightarrow \beta_i$. By the purity of
all states involved, the Stinespring unitary dilation must have
the form $\alpha_i \otimes \proj{0}\rightarrow \beta_i\otimes
\xi_i$ for some set $X=\{\xi_i\}$ of pure states. Then (with
suitable choices of kets) $G_A=G_X\circ G_B$. Since $\tr\,
\beta_i\beta_j\neq 0$ we have $[G_B]_{ij}\neq 0$ for all $i,j$ so
$G_X$ is uniquely determined i.e. $\{\xi_i\}$ is fixed up to
unitary equivalence and all such sets produce the same physical
transformation on the source space. $\qed$

Our ultimate goal would be to characterise transformability
$\{\rho_i\}\Rightarrow\{\sigma_i\}$ between general sets of mixed
states. The above proof technique (relating the matrix $M$
directly to the Stinespring form of the physical transformation)
reveals limitations of our approach for this general context. For
example if both sets of mixed states are on a system $X$ then the
Stinespring form of the physical transformation is
\[ U:\rho^X_i\otimes \proj{0}^E \rightarrow \xi_{i}^{XE}
\hspace{5mm} \mbox{with $ \sigma_i=\tr_E \,\xi_{i}^{XE}$.}
\] For pure target states $\sigma_i$ we must have the product form
$\xi_{i}^{XE}=\sigma_i^X\otimes \omega_i^E$ and we get the
Hadamard product structure for the corresponding Gram matrices.
However for mixed target states the $XE$ register will generally
be entangled and we lose the simple Hadamard product structure.
Another, perhaps more fundamental, problem is the lack of a
suitable generalisation of the notion of Gram matrix for a general
set of {\em mixed} states $\{\rho_i\}$. For example the matrix
constructed from $\tr\, \rho_i\rho_j$ and $\tr\,
\rho_1\rho_i\rho_j$ in eq. (\ref{gram1}) no longer characterises
$\{\rho_i\}$ up to unitary equivalence. Hence we need some new
kind of construction that also behaves well under partial traces.
However if the target set is pure (but the source set is general)
we still retain the product structure of the $XE$ register above
and the problem of $\{\rho_i\}\Rightarrow \{ \proj{b_i}\}$ can be
reduced to the fully pure state case as follows.

\begin{corollary}\label{cor3} For any set $\{\rho_i\}$
let $\rho_i=\sum_\nu p_{i\nu} \alpha_{i\nu}$ with
$\alpha_{i\nu}=\proj{a_{i\nu}}$ be any pure state decompositions
of the states and let $\{\beta_i\}$ be any set of pure states.
Then $\{\rho_i\}\Rightarrow \{\beta_i\}$ iff there is a physical
transformation $\ct:\alpha_{i\nu}\rightarrow \beta_i$ i.e. iff
$\{\alpha_{i\nu}\}\Rightarrow\{ \beta_{i\nu}\}$ where
$\beta_{i\nu}=\beta_i$ for all $\nu$.\end{corollary}

\noindent {\bf Proof}\,\,  Since $\beta_i$ is pure, any
transformation sending $\rho_i$ to $\beta_i$ must also send every
pure state in the range of $\rho_i$ to $\beta_i$ too, giving the
forward implication. The reverse implication is immediate by the
linearity of $\ct$. $\qed$

In the next section we will introduce a broader class of
transformations, multi-probabilistic state transformations, which
have implications for the more difficult problem of
$\{\rho_i\}\Rightarrow\{\sigma_i\}$ when the source set is pure
(and the target set may be mixed).

\section{Multi-probabilistic transformations}

Consider a family of pure source states, $\{\alpha_i\}$,
$i=1,\ldots,n$, and $m$ families of corresponding pure target
states, $\{\beta_i^j\}$, $i=1,\ldots,n$, $j=1,\ldots,m$, all being
states on the same Hilbert space ${\cal H}$ of finite dimension
$d$.
\par
We want to consider here the question whether there is a
transformation which, for each $i$, sends $\alpha_i$ with
probability $P_i^j$ to $\beta_i^j$, for an admissible probability
matrix $P_i^j$, i.e. a matrix of non--negative entries satisfying
 $\sum_j P_i^j\leq 1$ for all $i$. The value $P_i^0=1-\sum_j
P_i^j$ is the probability of failure of the process, given the
source state labelled $i$. The transformations we seek should in
fact not only perform these transformations with the conditional
probabilities $P_i^j$, but also output the value of
$j\in\{0,1,\ldots,m\}$ as a measurement result. Such
transformations are known in the literature as \emph{instruments}
\cite{davlew}.
\par
We call such a state map a \emph{multi--probabilistic
transformation}, denoted
$$\ct:\{\alpha_i\} \stackrel{P_i^j}{\longmapsto} \{\beta_i^j\}.$$
Note that even though we introduce this transformation only on
some pure states, it is generally a physical transformation on all
density operators on $\ch$.
\par
As a first observation, note that we can in fact demand $\sum_j
P_i^j=1$ for all $i$: we simply augment the operation $\ct$ by a
preparation of some arbitrary pure $\beta^0$, if the failure event
``$0$'' occurs. This we will assume from now on.
\par
As in section \ref{purestates} we introduce kets $\ket{a_i}$ and
$\ket{b^j_i}$ with $\alpha_i=\proj{a_i}$, $\beta^j_i=\proj{b_i^j}$
and corresponding Gram matrices $G_A$ and $G_{B^j}$ for the sets
$\{\alpha_i\}$ and $\{\beta^j_i\}$.

\begin{theorem}
  \label{thm:mp:st}
  A multi--probabilistic transformation
  $$\ct:\{\alpha_i\} \stackrel{P_i^j}{\longmapsto} \{\beta_i^j\}$$
  exists iff there are positive semidefinite matrices $\Pi^1,\ldots,\Pi^m$ with
  ${\rm diag}\,\Pi^j=(P_1^j,\ldots,P_n^j)$, such that
  \begin{equation}
    \label{eq:mp:st}
    G_A=\sum_j \Pi^j \circ G_{B^j}.
  \end{equation}
\end{theorem}

\noindent {\bf Proof}\,\,
  For the forward implication, assume that such a $\ct$ exists. By augmenting the Hilbert space
  ${\cal H}$ by a suitable ancillary space ${\cal E}$ and a measurement output
  space
  ${\cal J}={\rm span}\{\ket{j}:j=1,\ldots,m\}$, this transformation can be thought of
  as an isometry
  $$U:{\cal H}\longrightarrow{\cal H}\otimes{\cal J}\otimes{\cal E},$$
  followed by tracing out ${\cal E}$ and measuring $\{\ket{j}:j=1,\ldots,m\}$.
  To conform to the requirements, the image of $\ket{a_i}$ under $U$ must be
  of the form
  $$U\ket{a_i}=\sum_j \sqrt{P_i^j}\ket{b_i^j}\otimes\ket{j}\otimes\ket{\varphi_i^j}.$$
  Isometry requires preservation of the inner products, i.e. for all $i,k$,
  \begin{equation}
    \label{eq:isometry}
    \inner{a_i}{a_k}=
            \sum_j \sqrt{P_i^j
            P_k^j}\,\inner{\varphi_i^j}{\varphi_k^j}
                                \cdot\inner{b_i^j}{b_k^j},
  \end{equation}
  using that the $\ket{j}$ form an orthonormal basis in ${\cal J}$.
  Then $\Pi^j_{ik}=\sqrt{P_i^j P_k^j}\,\inner{\varphi_i^j}{\varphi_k^j}$ clearly
  defines semidefinite matrices with the correct diagonal, and
  satisfying eq.~(\ref{eq:mp:st}).
  \par
  Conversely, assume eq.~(\ref{eq:mp:st}). Then we can write
  $\Pi^j_{ik}=\sqrt{P_i^j P_k^j}\,\inner{\varphi_i^j}{\varphi_k^j}$, with
  normalised vectors $\ket{\varphi_i^j}$ (because a positive semidefinite matrix
  is always expressible as a Gram matrix of vectors whose lengths are given by
  the diagonal - an easy generalisation of property \ref{prop2} above).
  Thus clearly eq.~(\ref{eq:mp:st}) is satisfied, which means that
  that the sets $\{\ket{a_i}\}$ and
  $\left\{\sum_j \sqrt{P_i^j}\ket{b_i^j}\otimes\ket{j}\otimes\ket{\varphi_i^j}\right\}$
  are related by an isometry $U$. This, followed by tracing over ${\cal E}$ and
  measuring $\{\ket{j}:j=1,\ldots,m\}$, is the desired map $\ct$.
  $\qed$

\par
We remark that in case we choose not to demand $\sum_j P_i^j=1$
but only $\leq 1$, the criterion of the above theorem simply
becomes
\begin{equation}
  \label{eq:mp:st:alt}
  G_A \geq \sum_j \Pi^j \circ G_{B^j},
\end{equation}
which, for $m=1$ (i.e., only one set of possible target states)
reduces to (a generalisation of) Chefles' probabilistic state
transformation theorem in \cite{chefles2}.
\par\medskip
We demonstrate now how theorem \ref{thm:mp:st} can be applied to
the problem of $\{\rho_i\}\Rightarrow \{\sigma_i\}$ when the
source states $\rho_i$ are pure.

\begin{theorem}\label{puremix} Let $\{ \alpha_i\}$ and
$\{\sigma_i\}$ be sets of states with $\alpha_i$ pure. Then
$\{\alpha_i\}\Rightarrow \{\sigma_i\}$ iff  there exist pure state
decompositions $\sigma_i=\sum_j P^j_i \beta^j_i$
($\beta_i^j=\proj{b^j_i}$) and a multi-probabilistic
transformation $\ct: \{\alpha_i\} \stackrel{P_i^j}{\longmapsto}
\{\proj{b_i^j}\}$. \end{theorem}

\noindent {\bf Proof}\,\, Suppose a physical transformation maps
$\alpha_i$ to $\sigma_i$, which as usual can be written as the
composition of an isometry
$$U:{\cal H}\longrightarrow{\cal H}\otimes{\cal E},$$
involving an ancillary space ${\cal E}$, followed by a partial
trace over ${\cal E}$. Introducing an arbitrary rank--one
measurement POVM $(E_j)_{j=1,\ldots,m}$ in ${\cal E}$ (e.g., a
complete von Neumann measurement in some basis), we observe the
following:
\begin{equation*}\begin{split}
  \sigma_i &= \tr_{{\cal E}} U\alpha_iU^\dagger \\
         &= \sum_j \tr_{{\cal E}} \left( (\1\otimes E_j)U\alpha_iU^\dagger \right),
\end{split}\end{equation*}
because the $E_j$ sum to unity. Since the $\alpha_i$ are pure, the
summands in the second line are positive rank--one operators,
which can be written
$$\tr_{{\cal E}} \left( (\1\otimes E_j)U\alpha_iU^\dagger \right)
                                                 = P_i^j \proj{b_i^j},$$
with pure states $\ket{b_i^j}$ and probability distributions
$(P_i^1,\ldots,P_i^m)$.
\par
Thus the sequence of operations: the isometry $U$, measuring the
POVM $(E_j)$ and tracing out ${\cal E}$, implements a
multi--probabilistic transformation
\begin{equation}
  \label{eq:pure:mixed:st}
  \ct :\{\alpha_i\} \stackrel{P_i^j}{\longmapsto} \{\ket{b_i^j}\}
  \hspace{5mm} \mbox{
  with $\sum_j P_i^j \ketbra{b_i^j} = \sigma_i$ for all
  $i$.}
 \end{equation}
Conversely, the existence of such a multi--probabilistic
transformation obviously implies that the transformation has the
property $\ct :\{ \alpha_i\} \longmapsto \{\sigma_i\}$. $\qed$

Hence we have found necessary and sufficient conditions for the
problem of mapping pure source states to general target states.
However the result is more complicated than the cases of theorem
\ref{pure} and corollary \ref{cor3} since it involves an
existential quantifier over all pure state decompositions of the
target states. As such, it is more difficult to implement in
practice but we note that our conditions may be recast in the
formalism of semi-definite programming to facilitate their
implementation. More details will be given in an expanded version
of this paper. The general problem of finding useful tractable
conditions guaranteeing $\{\rho_i\}\Rightarrow\{\sigma_i\}$, where
both sets may contain mixed states, appears to be a rather
difficult question. Some further recent developments have been
announced in \cite{albnew}.

\noindent {\large\bf Acknowledgements}\\ The authors were
supported by the UK Engineering and Physical Sciences Research
Council. AC was also supported by a University of Hertfordshire
Postdoctoral Fellowship for part of this work. We are grateful to
Peter Alberti for comments and references to earlier relevant
works.

\end{document}